\documentclass[twocolumn,preprintnumbers,amsmath,amssymb]{revtex4}
\usepackage{graphicx}

\begin{document}

\title{Double injection in graphene p-i-n structures} 
\author{V. Ryzhii$^{1}$, I. Semenikhin$^{2}$, M. Ryzhii$^{3}$, D. Svintsov$^{2}$, V.~Vyurkov$^{2}$, 
A. Satou$^{1}$,  and T. Otsuji$^{1}$} 
\affiliation{
$^1$ Research Institute for Electrical Communication, Tohoku University, 
Sendai 980-8577, 
 Japan\\
 $^2$ Institute of Physics and Technology, Russian Academy of Sciences,
Moscow 117218, Russia\\
$^3$ Department  of Computer Science and Engineering,
University of Aizu,
Aizu-Wakamatsu 965-8580, Japan\\
}

\begin{abstract}
We study the processes of the electron and hole injection (double injection) into the  i-region
of graphene-layer and multiple graphene-layer p-i-n
structures at the forward bias voltages. The  hydrodynamic equations governing the electron and hole transport in graphene coupled with the two-dimensional Poisson equation are employed. Using analytical  and numerical solutions of the equations of the model, we calculate  the band edge profile, the spatial distributions of the quasi-Fermi energies,   carrier density  and velocity,
and the current-voltage characteristics. In particular, we demonstrated that the electron and hole collisions can strongly affect these distributions. The obtained results can be used for the realization and optimization of graphene-based injection terahertz and infrared lasers.
\end{abstract}

\maketitle
\newpage
\section{Introduction}
Due to unique properties,   graphene-layer (GL)  and multiple-graphene-layer (MGL) structures~\cite{1}
  with p-n and p-i-n junctions are considered as novel building blocks for
a variety of electron, terahertz, and optoelectronic devices.
Such junctions have been realized  using  both  chemical doping ~\cite{2,3,4,5,6} and  "electrical" doping
in the gated structures~\cite{7,8,9,10}. The formation of p-n and p-i-n junctions with the electrically induced p- and n-regions is possible even in MGL structures with rather large number of GLs~\cite{11}. 
Different devices  based on p-i-n junctions in
GL and MGL structures have been proposed recently. 
In particular, the GL and MGL structures with reversed biased p-i-n junctions
can be used in  infrared (IR) and terahertz (THz) detectors~\cite{12,13,14,15,16,17,18,19} and the tunneling transit-time THz oscillators~\cite{20,21}.
The GL and MGL p-i-n junctions under the forward bias can be the basis of IR and THz lasers
exploiting the interband population inversion~\cite{22,23,24,25} (see also recent experimental results,~\cite{26,27,28}), in which  the injection of electrons and holes from the n- and p-regions is utilized ~\cite{29,30} instead of optical pumping. 
A simplified model of the GL and MGL injection lasers with the p-i-n junctions
was considered recently~\cite{30}. The model in question assumes that the recombination of electrons and holes in the i-region and the leakage thermionic and tunneling currents at the p-i and i-n interfaces are relatively small. This situation can occur in the structures with sufficiently
short i-regions or at relatively low temperatures when the recombination associated with the emission of optical phonons and thermionic leakage are weakened. 
In such a case, different components of the current can be considered as  perturbations, and  
 the spatial distributions of the electric potential and the carrier density along the i-region are virtually uniform. 
 However, in the p-i-n structures with relatively long i-regions
and at the elevated voltages the spatial distribution of the potential can be rather nonuniform, particularly, near the edges of the i-region, i.e. near the p-i and i-n interfaces. In this case, the electric field in  the i-region can be sufficiently strong. Such an effect  can markedly influence the
density of the injected carriers, the conditions of the population inversion and current-voltage characteristics.  

In this paper, we develop a model for the GL and MGL forward biased p-i-n structures which accounts for relatively strong recombination 
and consider its effect on  the characteristics which can be important for realization of IR/THz injection lasers.
The problem of the electron and hole injection (double injection) in GL and MGL p-i-n structures 
is complicated by the two-dimensional geometry of the device and the features of the carrier transport properties. This makes necessary to use two-dimensional Poisson equation for the self-consistent electric potential around the i-region and to invoke the hydrodynamic equations for carrier transport
in GL and MGLs. Thus, the present model  is a substantial generalization of the model applied
in the previous paper~\cite{30}. To highlight the effects of strong recombination and the nonuniformity of the potential distributions, we, in contrast to Ref.~\cite{30,31}, disregard for simplicity the effects
of the carrier heating or cooling    and the effects associated with the nonequilibrium optical phonons. This can be justified by the fact that the injection of electrons and holes from the pertinent contacts does not bring a large energy to the electron-hole system in the i-region
unless the applied bias voltage is large. 

The paper is organized as follows. In Sec.~II, we present the  GL and MGL  p-i-n structures under consideration and 
the device mathematical model. Section III deals with
an analytical solutions of the equations of the model for a special  case. Here the general equations are  reduced to a simple equation for the quasi-Fermi energies of electrons and holes as  functions of the applied bias voltage.
In Sec.~IV, the equations of the model are solved numerically for fairly general cases.
The calculations in Secs.~III and IV provide  the spatial distributions of the band edges,
the quasi-Fermi energies and densities of electrons and holes, and their average (hydrodynamic) velocities. It is shown, that the obtained analytical
distributions match well with those obtained numerically, except very narrow regions near the p-i
and i-n interfaces.
In Sec.~V, the obtained dependences are used for the calculation of the current-voltage
 characteristics of the GL and MGL  p-i-n structures. 
Sec. VI deals with the  discussions of  some limitations of the model and possible consequences  of their lifting.
In Sec.~VII we draw the main conclusions.

\section{Model}

\subsection{The structures under consideration}

We consider the GL or MGL p-i-n structures with  either chemically or electrically doped p- and n-region (see, the sketches of the structures on the upper and lower panels in Fig.~1).
We assume that the p- and n-regions (called in the following p- and n-contacts)  in single- or multiple-GL structures are strongly doped,
so that the Fermi energy in each GL in these structures $\varepsilon_{Fc}$ and hence, the built-in voltage
$V_{bi}$ are  sufficiently large:
$$
\varepsilon_{Fc}  = eV_{bi}/2  \gg k_BT,
$$
where $T$ is the temperature and $k_B$ is the Boltzmann constant.
In the single-GL structures with the electrically induced p- and n-contact regions,
$$
\varepsilon_{Fc}  = eV_{bi}/2  = \frac{\hbar\,v_W}{2} \sqrt{\frac{\kappa\,V_g}{eW_g}}
$$
where $\hbar$ and $\kappa$ are the Planck constant and the dielectric constant, respectively,
$v_W \simeq 10^8$~cm/s is the characteristic velocity of electrons and holes in GLs,
$W_g$ is the gate layer thickness, and $V_g$ is the gate voltage ($V_p = - V_g$ and $V_n = V_g$ (see Fig.~1). Figure~2 shows the schematic views of the band profiles of a GL p-i-n structure
at the equilibrium [see Fig.~2(a)] and at the forward bias [see Fig.~2(b)]. The Fermi energies of the  holes and electrons in the contact p- and n-regions are assumed to be $\varepsilon_{Fc}$.
It is shown that at $V = 0$, the densities of electron and hole (thermal)  in the i-region 
are relatively small. But at the forward bias these density become rather large due to
the injection. The injection is limited by the hole charge near the p-i junction and
by the electron  charge near the i-n junction. Due to this, the Fermi energy of the injected
electrons and holes, $\varepsilon_{Fi}$, at the pertinent barriers is generally smaller than $\varepsilon_{Fc}$. However, at the bias voltages comparable with the build-in voltage $V_{bi}$, these Fermi energy can be close to each other. The qualitative band profiles of Fig.~2 are in line with the band profiles found from the self-consistent numerical calculation shown in Sec.~IV.

 Due to high   electron and hole densities in the i-region under the injection conditions, the electron and hole energy distribution 
are characterized by the Fermi functions with the quasi-Fermi (counted from the Dirac point)
$\varepsilon_{Fe}$ and $\varepsilon_{Fh}$, respectively, and the common effective temperature $T$,
which is equal to the lattice temperature. 

\begin{figure}[t]
\center{\includegraphics[width=6.0cm]{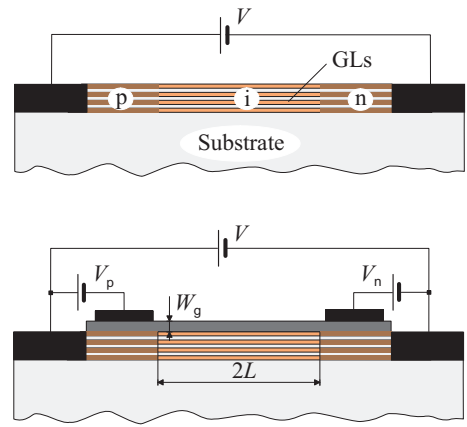}}
\caption{(Color online) Schematic view of the cross-sections
of MGL p-i-n structures   with chemically doped n- and p-contact regions (upper panel)
and  with such regions electrically induced by the side gate-voltages
$V_p = -V_g$ and $V_n = V_g >0$ (lower panel). 
}
\label{Fig1}
\end{figure}

\begin{figure}[t]
\center{\includegraphics[width=7.0cm]{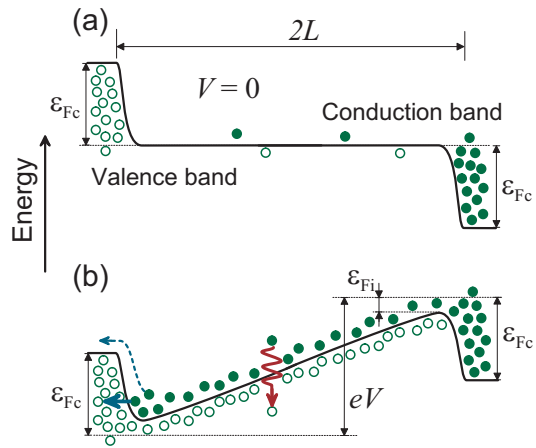}}
\caption{(Color online) Qualitative view of band profiles of a GL  p-i-n structure
(a) at $V = 0$ and (b) at forward bias $V > 0$.
Opaque and open circles correspond to electrons and holes, respectively. Wavy, straight, and dashed  arrows correspond to the recombination in the i-region (assisted by optical phonon emission), tunneling at the contact, and thermionic leakage to the contact, respectively 
.
}
\label{Fig2}
\end{figure}
\subsection{Equations of the model}

The model under consideration is based on the two-dimensional Poisson equation for the self-consistent electric potential and the set of hydrodynamic equations describing the electron and hole transport along the GL (or GLs) in the i-region.
The Poisson equation for the two-dimensional electric potential $\psi = \psi(x,z)$
is presented as
\begin{equation}\label{1}
\frac{\partial^2 \psi}{\partial x^2} + \frac{\partial^2 \psi}{\partial z^2}
= \frac{4\pi\,e}{\kappa} (\Sigma_e - \Sigma_d - \Sigma_h + \Sigma_a)\cdot \delta(z).
\end{equation}
where $e = |e|$ is the electron charge,  $\Sigma_e$ and $\Sigma_h$ are the net electron and hole sheet densities in the system (generally consisting of $K$ GLs),
$\Sigma_d = \Sigma_d(x)$ and $\Sigma_a = \Sigma_a(x)$ are the densities of donors and acceptors  in the  i-region (located primarily near the contact regions),  and $\delta(z)$ is the delta function reflecting the fact that the GL (MGL system) are located in a narrow layer near the plane  $z = 0$.
The axis $x$ is directed in this plane.

The transport of electrons and holes is governed by the following system of hydrodynamic equations~\cite{32}:

\begin{equation}\label{2}
\frac{d \Sigma_e u_e}{d x} = -R, \qquad \frac{d \Sigma_h u_h}{d x} = -R.
\end{equation}

\begin{equation}\label{3}
\frac{1}{M}\frac{d(e\varphi - \varepsilon_{Fe})}{dx} = \nu\,u_e + \nu_{eh}(u_e - u_h),
\end{equation}

\begin{equation}\label{4}
-\frac{1}{M}\frac{d(e\varphi + \varepsilon_{Fh})}{dx} = \nu\,u_h + \nu_{eh}(u_h - u_e).
\end{equation}

\begin{equation}\label{5}
\Sigma_e = \frac{2k_B^2T^2K}{\pi\hbar^2v_W^2}\int_0^{\infty}\frac{dyy}{[\exp(y - \varepsilon_{Fe}/k_BT) +1]},
\end{equation}
\begin{equation}\label{6}
\Sigma_h = \frac{2k_B^2T^2K}{\pi\hbar^2v_W^2}\int_0^{\infty}\frac{dyy}{[\exp(y - \varepsilon_{Fh}/k_BT) +1]}.
\end{equation}

Here $\varphi = \psi|_{z = 0}$ is the potential in the GL plane, 
$\varepsilon_{Fe}$, $\varepsilon_{Fh}$, $u_e$, and $u_h$ are the quasi-Fermi energies
and hydrodynamic velocities of electrons and holes, respectively, $\nu$ is the  collision frequency of electrons and holes with impurities and acoustic phonons, $\nu_{eh}$ is their collision frequency  with each other,
and $M$ is the fictitious mass, which at the Fermi energies of the same order of magnitude as
the temperature can be considered as a constant.
The recombination rate $R$ in the case of dominating optical phonon mechanism can be presented in the following simplified form (see, for instance, Refs.~\cite{30,31,33}:

$$
 R = \frac{K\Sigma_T}{\tau_R} \biggl[\frac{{(\cal N}_0 + 1)}{N_0}
 \exp\biggl(\frac{\varepsilon_{Fe} +\varepsilon_{Fh}  - \hbar\omega_0}{k_BT}\biggr) - 1\biggr] 
$$ 
 \begin{equation}\label{7}
 \simeq \frac{K\Sigma_T}{\tau_R} 
 \biggl[\exp\biggl(\frac{\varepsilon_{Fe} +\varepsilon_{Fh}}{k_BT}\biggr) - 1\biggr].
\end{equation}
Here $\Sigma_T = \pi\,k_B^2T^2/6\hbar^2v_W^2$ is the equilibrium density of electrons and holes in the i-region in one GL at the temperature $T$, $\tau_R = \Sigma_T/G_T$ is the characteristic time of electron-hole recombination associated with the emission of an optical phonon, 
 $G_T \propto {\cal N}_0$ is the rate of thermogeneration of the electron-hole pairs due to the absorption of
optical phonons in GL at equilibrium,   $\hbar\omega_0$ is the optical phonon energy,
and ${\cal N}_0 = [\exp(\hbar\omega_0/k_BT) - 1]^{-1}$.

\subsection{Boundary conditions}

The boundary conditions for the electron and hole velocities and Fermi energies are taken in the following form:

\begin{equation}\label{8}
u_e|_{x = - L} = - u_R, \qquad u_h|_{x = L} =u_R,
\end{equation}
and
\begin{equation}\label{9}
\varepsilon_{Fe}|_{x = L} = \varepsilon_{Fh}|_{x = - L} = \varepsilon_{Fc}/2 = eV_{bi}/2,
\end{equation}
where $2L$ is the spacing between p- and n-region (length of the i-region),
$u_R$ is the recombination velocity of electrons and holes  in narrow  space-charge regions adjacent to the p- and n-region, respectively,
due to the interband tunneling and due to the leakage over the barriers (edge recombination velocity)~\cite{7,34}, and $V$ is the applied bias voltage (see Fig.1) .
The quantity $\varepsilon_{Fc}$ is the Fermi energy at the contact regions.
The electric potential at the the contact regions is determined by the applied voltage $V$:

\begin{equation}\label{10}
(e\psi + \varepsilon_{Fh})|_{x \leq -L, z = 0} = \frac{eV}{2},
\end{equation}

 \begin{equation}\label{11}
\qquad (e\psi - \varepsilon_{Fe})|_{x \geq L, z = 0} = -\frac{eV}{2},
\end{equation}
where $V$ is the applied bias voltage. Combining Eqs.~(9) -  (11), one obtains

\begin{equation}\label{12}
\psi|_{x \leq -L, z = 0} =  \varphi|_{x = -L} = \frac{e(V - V_{bi})}{2},
\end{equation}
\begin{equation}\label{13}
\psi|_{x \geq L, z = 0} = \varphi|_{x = L} = - \frac{e(V - V_{bi})}{2}.
\end{equation}
The dependence $e\varphi = e\varphi(x)$ yields the coordinate dependence of the Dirac point or the band edge profile
(with respect to its value in the i-region center, $x = 0$).

\subsection{Dimensionless equations and boundary conditions}

To single out the characteristic parameters of the problem we introduce the following
 dimensionless variables: $\Psi = e\psi/k_BT$, $\Phi = e\varphi/T$,
$\mu_{e} = \varepsilon_{Fe}/T$, $\mu_{h} = \varepsilon_{Fh}/k_BT$, $\mu_c = 
\varepsilon_{Fc}/k_BT$,
 $\sigma_e = \Sigma_e/\Sigma_T$,
$\sigma_h = \Sigma_h/\Sigma_T$, $\sigma_d = \Sigma_d/\Sigma_T$, $\sigma_a = \Sigma_a/\Sigma_T$,  $U_e = u_e\tau_R/L$, $U_h = u_h\tau_R/L$, $U_R = u_R\tau_R/L$,
$\xi = x/L$, and $\zeta = z/L$. Using these variables,  
Eqs. ~(1) (6) are presented as

\begin{equation}\label{14}
\frac{\partial^2 \Psi}{\partial \xi^2} + \frac{\partial^2 \Psi}{\partial \zeta^2}
= 4\pi Q(\sigma_e - \sigma_d - \sigma_h + \sigma_a)\cdot \delta(\zeta).
\end{equation}

\begin{equation}\label{15}
\frac{d\sigma_eU_e}{d \xi} =
1 - \exp(\mu_e + \mu_h),
\end{equation}
\begin{equation}\label{16}
\frac{d\sigma_hU_h}{d \xi} =
1 - \exp(\mu_e + \mu_h),
\end{equation}

\begin{equation}\label{17}
\frac{d(\Phi - \mu_e)}{d\xi} = 
q\biggl[U_e\biggl(\frac{\nu}{\nu_{eh}}\biggr) +U_e - U_h\biggr],
\end{equation}

\begin{equation}\label{18}
-\frac{d(\Phi + \mu_h)}{d\xi} = 
q\biggl[U_h\biggl(\frac{\nu}{\nu_{eh}}\biggr) +U_h - U_e\biggr],
\end{equation}
\begin{equation}\label{19}
\sigma_e = \frac{12}{\pi^2}\int_0^{\infty}\frac{dyy}{[\exp(y - \mu_e) +1]},
\end{equation}

\begin{equation}\label{20}
\sigma_h = \frac{12}{\pi^2}\int_0^{\infty}\frac{dyy}{[\exp(y - \mu_h) +1]}.
\end{equation}
Here $Q$ and $q$ are the "electrostatic" and "diffusion" parameters given, respectively, 
by the following formulas:
$$
Q = K\frac{\pi\,e^2Lk_BT}{6\kappa\hbar^2v_W^2}, \qquad q = \frac{M\nu_{eh}L^2}{\tau_Rk_BT}.
$$
Assuming, $L = 1 - 5~\mu$m, $K = 1 - 2$, $\kappa = 4$, $T = 300$~K, $\varepsilon_{F0} = 100$~meV, $\nu_{eh} = 10^{13}$~s$^{-1}$,
$\tau_R = 10^{-9}$~s, we obtain $Q = 12.5 - 125$,  $q = 0.04 - 1.0$.
The same values of $q$ one obtains if $\nu_{eh} = 10^{12}$s~$^{-1}$ and $\tau_R = 10^{-10}$~s.
%
%
The parameters $Q$ and $q$ can also be presented as $4\pi\,Q \sim L/r_S^T$ and $q \sim (L^2/D\tau_R) \simeq (L/L_D)^2$, where $r_S^T = (\kappa\hbar^2v_W^2/4e^2T)$ 
and  $L_D = \sqrt{D_i \tau_R}$ are the characteristic screening  and the diffusion lengths in the i-region (in equilibrium when $\mu_i = 0$), respectively,
and $D= v_W^2/2\nu_{eh}$  is  the diffusion coefficient.

The boundary conditions
for the set of the dimensionless Eqs.~ (14) - (20) are given by
\begin{equation}\label{21}
\mu_e|_{\xi = 1} = \mu_h|_{\xi = -1} = \mu_c = eV_{bi}/2k_BT, 
\end{equation}
\begin{equation}\label{22}
U_e|_{\xi = -1} = - U_R, \qquad U_h|_{\xi = 1} =  U_R,
\end{equation}
\begin{equation}\label{23}
\Psi|_{\xi \leq -1, \zeta = 0} =  \Phi|_{\xi = -1 } = eV/2k_BT - \mu_c,
\end{equation}
\begin{equation}\label{24}
\Psi|_{\xi \geq 1, \zeta = 0} = \Phi|_{\xi = 1} =  -eV/2kBT  + \mu_c.
\end{equation}

In the following we focus our consideration mainly on the structures with very abrupt p-i and i-p
junction neglecting terms $\sigma_d$ and $\sigma_a$ in Eq.~(14). The effect of smearing of these junctions
resulting in $\sigma_d \neq 0$ and $\sigma_a \neq 0$ will be briefly discussed in Sec.~VI.

\section{Injection characteristics (Analytical approach for a special case)}

\begin{figure}[t]
\center{\includegraphics[width=6.0cm]{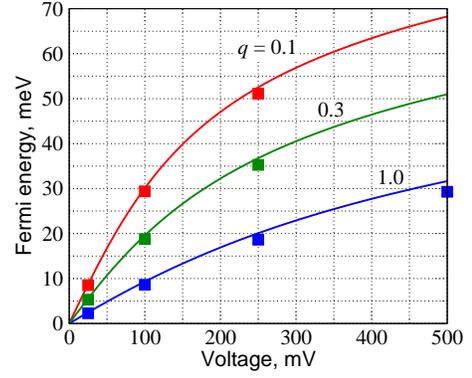}}
\caption{(Color online) Fermi energy in the i-region $\varepsilon_F$
as a function of the bias voltage $V$ for different values of parameter $q$.
Opaque squares correspond to electron (hole) Fermi energies at $x = 0$ calculated numerically in Sec.~IV.
}
\label{Fig3}
\end{figure}

Consider the special case in which $\nu/\nu_{eh}, U_R \ll 1$.
When $Q \gg 1$, Eq.~(14) is actually a partial differential equation with a small parameter
($Q^{-1} \ll 1$) at highest derivatives. Usually the net solution of such equations can be combined
from their solutions disregarding the left side (valid in a wide range of independent variables) and the solution near the edges (which are affected by the boundary conditions) in the regions of width $\eta \sim Q^{-1} \sim r_S^T/L \ll 1$,
where the derivatives are large, provided a proper matching of these solutions~\cite{35}.
Hence, since parameter $Q \gg 1$,  for  a wide  portion of the i-region one can assume that $\sigma_e \simeq \sigma_h \simeq \sigma_i = const$
and, consequently, $\mu_e \simeq \mu_h = \mu_i$.

In this case, Eqs.~(15) and (16) yield

\begin{equation}\label{25}
U_e = -\frac{(e^{\displaystyle 2\mu_i} - 1)}{\sigma_i}(\xi + 1),
\end{equation}
\begin{equation}\label{26}
U_h = - \frac{(e^{\displaystyle 2\mu_i} - 1)}{\sigma_i}(\xi - 1),
\end{equation}

It is instructive that the condition $\mu_e = \mu_h = \mu_i \simeq const$, assumed above,
is fulfilled only if the collisions with impurities and acoustic phonons are disregarded.
Therefore, in this section we assume that $\nu \ll \nu_{eh}$. Considering this, we find

\begin{equation}\label{27}
\Phi = -2q\frac{(e^{\displaystyle 2\mu_i} - 1)}{\sigma_i}\xi.
\end{equation}
At the points $\xi = \pm (1 - \eta) \simeq \pm 1$, one can use the simplified matching conditions [see Fig.~2(b)] and obtain
the following equation for $\mu_i$:

\begin{equation}\label{28}
\mu_i + 2q\frac{(e^{\displaystyle 2\mu_i} - 1)}{\sigma_i}= \frac{eV}{2k_BT}
\end{equation}
or
\begin{equation}\label{29}
\mu_i  +  \frac{2q\pi^2(e^{\displaystyle 2\mu_i} - 1)}
{12\displaystyle \int_0^{\infty}\frac{dyy}{[\exp(y - \mu_i) +1]}}
 =
\frac{eV}{2k_BT}.
\end{equation}

If $eV < 2k_BT$, one can expect  that $2\mu_i < 1$, so that $\sigma_i \simeq 1$, and
Eq.~(29) yields

\begin{equation}\label{30}
\mu_i \simeq \frac{eV}{2k_BT(1 + 4q)}
\end{equation}
At $eV > 2T$, $2\mu_i > 1$ (the electron and hole components are  degenerate),
from Eq.~(29)

\begin{equation}\label{31}
\mu_i + \frac{\pi^2q}{3\mu_i^2}e^{\displaystyle 2\mu_i} = \frac{eV}{2k_BT}.
\end{equation}

If formally $q \rightarrow 0$ (very long diffusion length or near ballistic transport of electrons and holes in the i-region), one can see from Eqs.~(29) - (31) that
$\mu_i$ tends to $eV/2T$, i.e.,
the Fermi energies of electrons and holes tend to $eV/2$ (as in the previous paper~\cite{30}).

Figure~3  shows the voltage dependences of the Fermi energy, $\varepsilon_{Fe}  = T\mu_i \simeq  \varepsilon_{Fh}$, of electrons and holes in the main part of the i-region calculated using Eq.~(29) for different values of parameter $q \propto (L/L_D)^2$. 
The dependences shown in Fig.~3 imply that the recombination leads to the natural decrease in the  Fermi energies and the densities of electrons and holes in the i-region. These quantities increase with increasing
applied voltage $V$, but such an increase is a slower function of $V$ (logarithmic) than
the linear one.
The results of the calculations based on Eq.~(29) (i.e.,  based on a simplified model valid for weak   electron and hole collisions with impurities and acoustic phonons
and  the edge recombination)   practically coincide with the results
of numerical calculations involving a rigorous model (shown by the markers in Fig.~3).
The results of these calculations  are 
considered in the next Section.

\section{Numerical results}

Finding of analytical solutions of Eq.~(14) coupled with Eqs.~(15) - (20) in the near-edge region and 
their matching with the smooth solutions in the main portion of the i-region
is a difficult problem
due to the nonlinearity of these system of equations and its integro-differential nature.
In principle, what could such solutions  provide is an  information  on the width of the near-edge regions.
The latter is not so important, because the main characteristics of the structures under consideration are determined by the electron-hole system in the structure bulk (in the main portion of the i-region). However, the net solution is particularly interesting to verify the results of
the analytical model considered in Sec.~III. For this purpose, we solved Eqs.~(14) - (20) numerically by iterations.

Due to large values of parameter $Q$, one could expect very sharp behavior of $\Psi$ and $\Phi$
near the edges of the i-region, so that strongly nonuniform mesh was used.
The potential $\Phi$ was found by reducung the Poisson equation (14) to the following:

$$
\Phi(\xi) = -\frac{V/T - 2\mu_c}{\pi} \sin^{-1}\xi 
$$
\begin{equation}\label{32}
+ \frac{4\pi\,Q}{k}\int_{-1}^{1}d\xi^{\prime} G(\xi, \xi^{\prime})[\sigma_h(\xi^{\prime})
- \sigma_e(\xi^{\prime})],
\end{equation}
where
$$
G(\xi, \xi^{\prime}) = \frac{1}{4\pi}\ln\biggl\{\frac{1 + \cos[\sin^{-1}(\xi) +\sin^{-1}(\xi^{\prime})]}
{1 - \cos[\sin^{-1}(\xi) -\sin^{-1}(\xi^{\prime})]}\biggr\}
$$
is the Green function, which corresponds to the geometry and the boundary conditions
under consideration.
%

\begin{figure}[t]
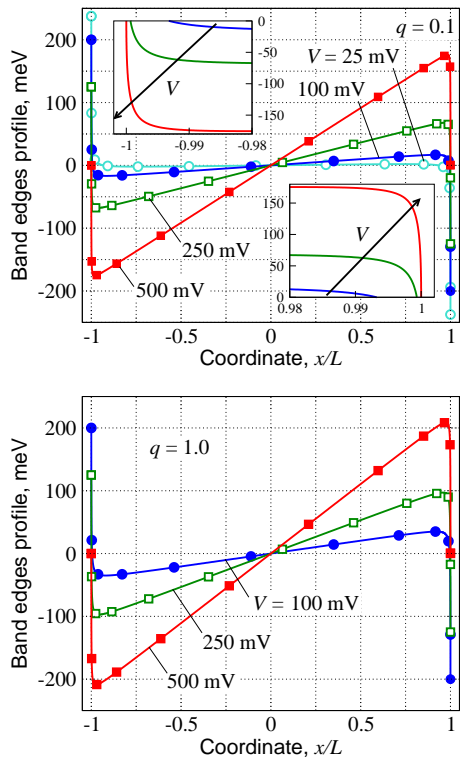

\center{\includegraphics[width=6.0cm]{DOUBLE_INJECTION_F4a.eps}}
\center{\includegraphics[width=6.0cm]{DOUBLE_INJECTION_F4b.eps}}
\caption{(Color online)  Conduction band bottom (valence band top) profiles 
in the i-region calculated for different bias voltages $V$ at  $q = 0.1$ (upper panel) and  $q = 1.0$ (lower panel) at 
$Q = 100$. The extreme left and right markers (at $x/L = \mp 1$) show the positions of the Dirac point at  p-i and i-n junctions. Insets show detailed behavior in close vicinities near the i-region edges.
}
\label{Fig4}
\end{figure}

\begin{figure}[t]
\center{\includegraphics[width=6.0cm]{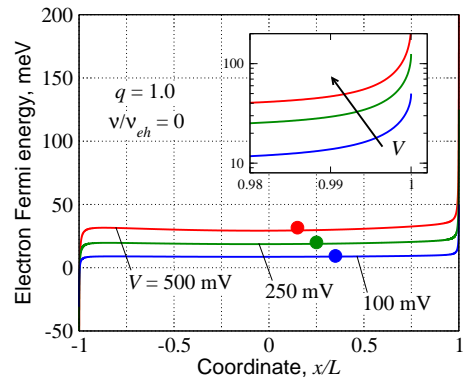}}\\
\caption{(Color online) Spatial distributions  of the electron Fermi energy
$\varepsilon_{Fe}$
 at $q = 1$ and $Q = 100$ calculated for different values of  bias voltage $V$ 
  at $U_R = 0$. Inset 
 shows dependences near
 the n-contact. Opaque circles  correspond to data obtained using analytical model.
}
\label{Fig5}
\end{figure}

\begin{figure}[t]
\center{\includegraphics[width=6.0cm]{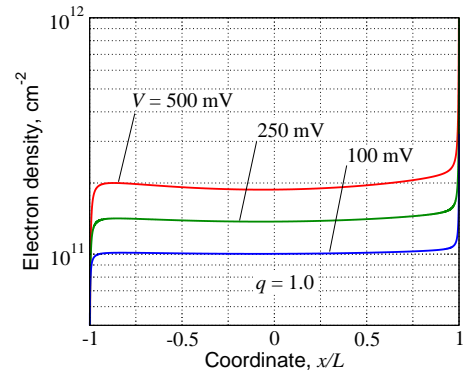}}
\caption{(Color online)  Spatial distributions  of 
 electron density $\Sigma_e$ at different bias voltages $V$ ($\nu/\nu_{eh} = 0$ and $U_R = 0$).
}
\label{Fig6}
\end{figure}
\begin{figure}[t]
\center{\includegraphics[width=6.0cm]{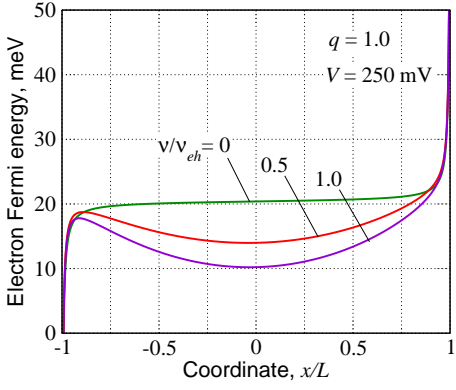}}
\caption{(Color online) Spatial distributions  of the electron Fermi energy
$\varepsilon_{Fe}$
 at $q = 1$ and $Q = 100$ calculated for different values of    ratio $\nu/\nu_{eh}$  and $V = 250$~mV  at $U_R = 0$.
}
\label{Fig7}
\end{figure}

\begin{figure}[t]
\center{\includegraphics[width=6.0cm]{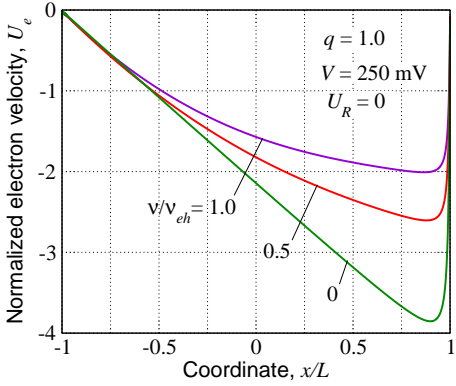}}
\caption{(Color online)  Coordinate dependences of normalized
 electron velocity $U_e$ calculated for $q = 1.0$ and $V = 250$~mV at different values of  $\nu/\nu_{eh}$.
}
\label{Fig8}
\end{figure}

\begin{figure*}[t]
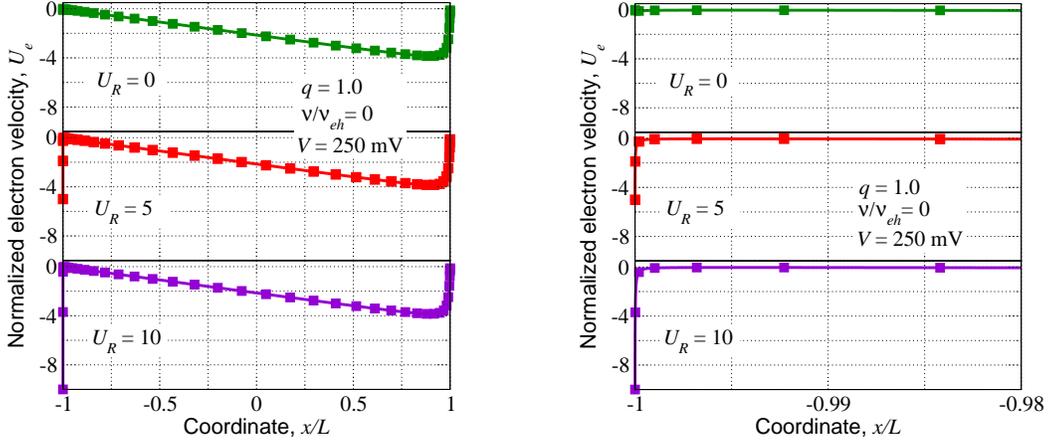

\center{\includegraphics[width=6.0cm]{DOUBLE_INJECTION_F9a.eps}\hspace{1.5cm}
\includegraphics[width=6.2cm]{DOUBLE_INJECTION_F9b.eps}}
\caption{(Color online)  The normalized 
 electron velocity $U_e$ versus coordinate $x$ for $q = 1.0$ and $V = 250$~mV at different values of  normalized edge recombination velocity $U_R$ (left panel). Right panel shows details of the same dependences in close vicinity of the p-i interface.
}
\label{Fig9}
\end{figure*}

Figure~4 shows the profiles of the band edges, i.e.,the conduction band bottom (valence band top) in the i-region  ($-L \leq x \leq L$)
obtained from numerical solutions of Eqs.~ (14) - (20) for different values of the bias voltage $V$ and parameter $q$  at $Q = 100$ (and $\nu/\nu_{eh}, U_R \ll 1$ as in Sec.~III).
As follows from Fig.~4, an increase in the bias voltage $V$ leads to the rise of the barrier
for the electrons injected from the n-region and the holes from the p-region.
These barriers are formed due to the electron and hole self-consistent surface charges localized in very narrow region (of the normalized width $\eta \sim Q^{-1} = 0.01$)
 in the i-region near its edges. The electric field near the edges changes the sign.
In the main portion of the i-region, the electric field is almost constant. Its value increases
with increasing parameter $q$, i.e., with reinforcement of the recombination.
Figures~5 and 6 demonstrate the spatial distributions of the electron Fermi energy $\varepsilon_{Fe}(x)$ and the electron density  $\Sigma_e(x)$ in 
the i-region for $q = 1.0$
at different bias voltages.
The hole Fermi energy and the 
 hole density are equal, respectively, to $\varepsilon_{Fh}(x) = \varepsilon_{Fe}(-x)$ and
 $\Sigma_h(x) = \Sigma_e(-x)$.

One can see from Figs.~5 and 6 that the electron Fermi energy and the electron density steeply
drop in fairly narrow region adjacent to the p-region. However, they remain virtually constant
in the main portion of the i-region if the ratio  of the  collisions frequency of electrons and holes with impurities and acoustic phonons and the frequency of electron-hole collisions
$\nu/\nu_{eh} \ll 1$. 
In the latter case, the values of these quantities across almost the whole i-region
coincide with those obtained analytically above with a high accuracy (see the markers in Figs.~3 and 5). This justifies the simplified approach used in the analytical model for the case $\nu/\nu_{eh}, U_R \ll 1$ developed in Sec.~III. Nevertheless, at not too small values of parameter  $\nu/\nu_{eh}  $,  the coordinate dependences of the electron (and hole) Fermi energy  exhibit a pronounced sag (see  Fig.~7). Similar sag  was observed  in the coordinate dependences
of the electron and hole densities (not shown). 
As seen from Fig.~8, an increase in the ratio  $\nu/\nu_{eh}$ leads also to a pronounced modification of the coordinate dependences of the electron (hole) velocity: the absolute value of the electron
velocity markedly decreases with increasing  $\nu/\nu_{eh}$.
However, the changes in the band edge profiles and, hence the electric field in the main portion of the i-region are insignificant.

Above, we have considered the cases of the p-i-n structures with sufficiently long i-region,
so that the normalized edge recombination velocity  $U_R \ll 1$. In such cases, in accordance with 
the boundary conditions $U_e$ (and $U_h)$ tends to zero at the pertinent contact region (see Fig.~8).

The calculations of the spatial distributions of the band edge profiles, Fermi energies, and
carrier concentrations showed that the variation of the edge recombination velocity $U_R$
(at least in the range from zero to ten) does not lead to any pronounced distinction even in the
near-edge regions. 
In all cases considered, as seen from Fig.~9, the absolute value of $U_e$ steeply increases in a narrow region near the n-contact and the gradually drops
(virtually linearly) across the main portion of the i-region. 
However, $U_e$ (and, hence, the electron hydrodynamic velocity $u_e$),
being virtually independent of $U_R$ from the n-region to the near-edge region adjacent to the p-i interface, strongly depends on $U_R$ in the latter region [see Fig.~9 (both left and right panels)].
 The hole velocity exhibits the same behavior with $U_h(x) = - U_e(-x)$.

Due to rather short near-contact regions, 
the edge recombination does not affect
substantially the integral characteristics of the p-i-n structures
under consideration, in particular total number of the injected electrons and holes
and their net recombination rate
at least
in the practical p-i-n structures
with $Q \gg 1$ and not too large $U_R$ (i.e., not too short i-regions).

\section{Current-voltage characteristics}

The current (associated with the recombination in the i-region) can be calculated
as

\begin{equation}\label{33}
J = e\int_{-L}^LdxR = \frac{eKL\Sigma_T}{\tau_R}\int_{-1}^1d\xi\biggl[\displaystyle e^{\displaystyle (\mu_e + \mu_h)} - 1\biggr].
\end{equation}

Using the obtained analytical formulas derived above, in which $\mu_e(x) \simeq \mu_h(x) =\mu_i \simeq const$, we find

\begin{equation}\label{34}
J = \frac{2eKL\Sigma_T}{\tau_R}(\displaystyle e^{\displaystyle 2\mu_i} - 1)
\end{equation}

or

\begin{equation}\label{35}
J = KJ_0(\displaystyle e^{\displaystyle 2\mu_i} - 1),
\end{equation}
where
\begin{equation}\label{36}
J_0 =\frac{\pi\,eLk_B^2T^2}{3\hbar^2v_W^2\tau_R}
\end{equation}
Setting $L = 1 - 5~\mu$m and $\tau_R = 10^{-10} - 10^{-9}$~s at $T = 300$~K, from Eq.~(34) one obtains
$J_0 \simeq (3 - 150)\times10^{-3}$~A/cm.

\begin{figure}[t]
\center{\includegraphics[width=6.0cm]{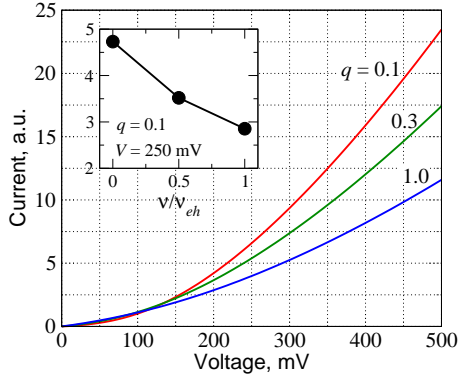}}\\
\caption{(Color online)  Current-voltage characteristics for different values of parameter $q$. Inset shows the current as a function of  parameter  $\nu/\nu_{eh}$ calculated
at $V = 250$~mV for $q = 1.0$.
}
\label{Fig10}
\end{figure}

The current-voltage characteristics obtained
for different parameters $q$ (using Eqs.~(29) and (35) at $\nu/\nu_{eh} \ll 1$) are
shown in Fig.~10. 
As follows from Fig.~10, the injection current superlinearly increases
with increasing bias voltage $V$. The steepness of such current-voltage characteristics decreases
with increasing  parameters $q$. 
This is because a larger $q$ corresponds in part to larger $\nu_{eh}$, i.e,  to
a larger resistance. In the limit $q \rightarrow 0$, the current-voltage
 dependence
approaches to $J \propto \exp(eV/2T)$. Thus, at the transition from near ballistic electron ($q \ll 1) $ and hole transport to collision dominated transport ($q \sim 1$),
the current-voltage characteristics transform from  exponential to  superlinear ones.

However, at finite values of parameter $\nu/\nu_{eh}$,
$\mu_i$ can be a prononced function of the coordinate as seen from Fig. ~7.
This results in a dependence of the current on the parameter in question.
Inset in Fig.~10 shows the dependence of the current on parameter $\nu/\nu_{eh}$
at fixed bias voltage.
A decrease in  the current   with increasing
parameter $\nu/\nu_{eh}$ is associated with a sag in the spatial dependence of the Fermi energy
(see Fig.~7), which gives rise to
weaker recombination in an essential portion of the i-region and, therefore, lower recombination (injection) current.

\section{Discussion}

The rigorous calculation of the edge recombination current, particularly, its tunneling component and parameter $U_R$
requires special consideration. Here we restrict  ourselves to rough estimates based of a phenomenological treatment. The characteristic velocity, $u_R^{tunn}$, of the edge recombination due to
tunneling of electrons through the barrier at the p-i-interface (and holes through the barrier at the i-n-interface) can be estimated as $u_R^{tunn} = D^{tunn}v_W/\pi$. Here $D^{tunn}$ is the effective  barrier tunneling transparency, determined, first of all, by the shape of the barrier. The appearance of factor $1/\pi$ is associated with the spread in the directions of electron velocities in the GL plane.  In the case of very sharp barrier, $D^{tunn} \simeq 1$. But practically $D^{tunn}$  is less than unity, although due to the gapless energy spectrum of carriers in GLs and MGLs one can not expect that 
it is very small. 
The edge recombination velocity associated with the thermionic current of electrons (holes) over the pertinent barrier is proportional to $D_{th} \propto exp(-\mu_c)$, so that
$u_R^{th} \propto (v_W/\pi)\exp(-\mu_c) $. 
The factor $\exp(-\mu_c)$ is very small when the contact n- and p-regions
are sufficiently doped ($\mu_c \gg 1$),  hence, $U_R^{th} \ll U_R^{tunn}$. In the case of the electrically induced p- and n-regions
in MGL structures, the condition $\mu_c \gg 1$ can impose certain limitation on the number of GL $K$~\cite{11}.  As a result, parameter $U_R$ can be estimated as 
$U_R \simeq u_R^{tunn}\tau_R/L
= D^{tunn}v_W\tau_R/\pi\,L$.
At $\tau_R = 10^{-10}$~s, $L = 5~\mu$m, and $D^{tunn} = 0.5$ we find that $U_R \simeq 3$.
As shown above, the edge recombination even at relatively large parameter $U_R$ markedly influences the  carrier densities and their Fermi energies only in immediate vicinity of
the p- and i-n interfaces (see Fig.~9). To find  conditions when the surface (tunneling) recombination can be disregarded in comparison to the recombination assisted by optical phonons,
we compare the contributions of this recombination to the net current.
Taking into account  Eq.~(35) and considering that the edge recombination tunneling current
$J^{tunn} \simeq ev_WD^{tunn}\Sigma_T$, we obtain

\begin{equation}\label{37}
\frac{J}{J^{tunn}} \gtrsim \frac{2\pi\,L}{D^{tunn}v_W\tau_R}(\displaystyle e^{\displaystyle 2\mu_i} - 1)
\simeq \frac{2}{U_R^{tunn}}\displaystyle e^{\displaystyle 2\mu_i}.
\end{equation}
Equation~(37) yields the following condition
when the edge recombination is provides relatively small contribution to the injection current:
\begin{equation}\label{38}
L > \frac{D^{tunn}v_W\tau_R}{2\pi}\displaystyle e^{\displaystyle -2\mu_i}.
\end{equation}
Assuming that $\tau_R = 10^{-10}$~s, $D^{tunn} = 0.5$, and $\mu_i = 1.0 - 1.4$ ($\varepsilon_{Fi} = 25 - 35$~meV), from inequality (36) we obtain
$2L > (1 - 2)~\mu$m. As follows from Eq.~(38), the role of the edge recombination steeply drops when
$\mu_i$, i.e., the bias voltage $V$ increase. This is because the recombination rate  $R$, associated with optical phonons, rapidly rises with increasing $\mu_i$.
If $\tau_R = (10^{-10} - 10^{-9})$~s and
$\mu_i = 2.5$ ($\varepsilon_{Fi} = 62.5$~meV), one can arrive at the following condition: $2L > 0.1 - 1~\mu$m.


\begin{figure}[t]
\center{\includegraphics[width=6.0cm]{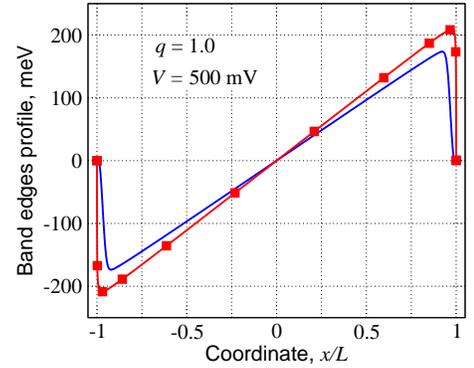}}\\
\caption{(Color online)  Comparison of  band
edge profiles in structures with abrupt (solid line with markers, the same as in Fig.~4)
and with smeared   p-i and n-i junctions (without markers). 
at $V = 500$~mV for $q = 1.0$. 
}
\label{Fig11}
\end{figure}

 As demonstrated in Sec.~IV, the spatial variations of the band edge profile, density of carriers and their Fermi energies are very sharp near the p-i and i-n interfaces. This is owing to large values of parameter
$Q$ and the assumption of abrupt doping profiles.
In this case, the normalized width of the transition region is $\eta \sim Q^{-1} \sim r_S^T/L \ll 1$.
 In real GL and MGL  p-i-n structures, the boundaries between p-, i-, and n-region are somewhat smeared (with the characteristic length $l \ll L$). This can be associated with the specifics of chemical doping or with the fringing effects. In the latter case, the length of the intermediate region is about the thickness of the gate layer $l \sim W_g$  (see Fig. 1). In both cases,
this can result in a smoothing of the dependences in near-contact regions
in comparison with those shown in Figs.~4 - 9 and in  a marked decrease
in the electric field in the  p-i and i-p junctions and in the edge tunneling current. 
Figure 11 demonstrates examples of the band edge profiles calculated for the abrupt p-i and i-n junctions and
for that with smeared junction. In the latter case, it was assumed that the acceptor and donor  densities varied exponentially with the characteristic length $l = 0.03L$: $\sigma_d \propto exp[-(\xi - 1)^2/\eta^2]$ and $\sigma_d \propto exp[-(\xi + 1)^2/\eta^2]$,
with $\eta = l/L = 0.03$,
$\Sigma_c =5.6\times 10^{12}$~cm$^{-2}$ ($\sigma_c = 62.8$, $\mu_c \simeq 10$). 
As seen from Fig.~11, an increase in smearing of the dopant distributions leads to a natural
increase in the widths of the transition regions and a decrease (relatively small)  in the electrical field in the i-region.

At moderate number of GLs $K$ in the p-i-n structures under consideration,
the net thickness of the latter $Kd$, where $d \sim 0.35$~nm is the spacing between GLs,
is smaller than all other sizes. In such a case, the localization of the space charges of electrons and holes in the i-region in the $z$-direction can be described by the delta function  $\delta(z) $ as in Eq.~(1).
In the case of large $K$,  the delta function should be replaced by a function with a finite localization width. This also should give rise to an extra smearing of the spatial distributions in the $x$-direction.
The pertinent limitation on the value $K$ can be presented as follows:
$Kd \ll r_S, l \ll  L$, where $r_S = [r_S^T/\ln(1 + \displaystyle e^{\mu_i})]$ is the screening length.
At $r_S \lesssim l \ll L$, assuming that $\kappa= 4$ and  $\mu_i \gtrsim 1$, one can obtain $r_S \simeq 10$~nm and arrive at the following condition $K \ll 30 $.
An increase in $K$ results in  an increase in parameters $Q$ and, hence, in some modification  
of the spatial distributions in near contact region, although the latter weakly affects these distributions in the main portion of the i-region. As a result, the injection current increases proportionally to $K$ (see Eq.~(35)).

We disregarded possible  heating (cooling) of the electron-hole system in the i-region.
Under  large bias voltages the electric field in this region can be fairly strong.
In this case, the dependence of the drift electron and hole velocities on the electric field might be essential.
However,  relatively strong interaction of electrons and holes with optical phonons having fairly large energy
prevents substantial heating~\cite{36}  in the situations considered in the previous sections. Moreover,
in the essentially bipolar system the recombination processes with the emission of large energy optical phonons
provide a marked cooling of the system, so that its effective temperature can become even lower than the lattice temperature~\cite{30,31}.

The above consideration was focused on the double injection in GL and MG p-in structures at
the room temperatures. The obtained results are also applicable at somewhat lower temperatures
if the recombination is primarily associated with  optical phonons. However, at lower temperatures,
other recombination mechanisms can become more  crucial, for instance, the  Auger recombination mediated by  scattering mechanisms on impurities and  acoustic phonons and  due to trigonal warping of the GL band structure, which provide
additional momentum and enables transitions to a broader range of final states~\cite{37,38,39}, i.e.,
the indirect Auger recombination~\cite{40}
(see also Refs.~\cite{41,42}), 
acoustic phonon recombination, invoking the momentum transfer due to disorder~\cite{43}, radiative recombination~\cite{44},  
tunneling recombination in the electron-hole puddles~\cite{45} and  edge tunneling  recombination (see Ref.~\cite{34} and Sec.~IV).  Relatively weak role of the optical phonon assisted processes at low temperatures can result in pronounced hot carrier effects~(see, for instance, Refs.~\cite{46,47}).
However, these effects are beyond the scope of our paper and require special consideration.

\section{Conclusions}

\vspace*{-3mm}
We  studied  theoretically the double electron-hole injection in
GL and MGL  p-i-n
structures at the forward bias voltages using the developed device model.
The mathematical part of the model was based on a system of
hydrodynamic equations governing the electron and hole transport in GLs  coupled with the two-dimensional Poisson equation. 
The model is characterized primarily by the following two parameters: the electrostatic parameter $Q$ and the diffusion
parameter $q$. Large value of parameter $Q$ in practical structures results in 
the formation of very narrow edge regions near p-i and i-n interfaces, in which all physical quantities vary sharply, and the formation of   a wide region, which is stretched across   the main portion of the i-region, 
where all spatial dependences are rather smooth.
Using analytical  and numerical solutions of the equations of the model, we calculated  the band edge profiles, the spatial distributions of the quasi-Fermi energies,   carrier density  and velocity,
and the current-voltage characteristics. It was demonstrated that the electron-hole collisions and the collisions of electrons and holes with impurities and acoustic phonons 
 can strongly affect the characteristics of the p-i-n structures. In particular, such collisions result in a pronounced
 lowering of the injection efficiency (weaker dependences of the electron and hole Fermi energies and their densities on the bias voltage) and the modification of the current-voltage characteristics
 from an exponential  to superlinear characteristics
 (at collision-dominated transport). It is also shown that for the case of relatively perfect
 GL and MGL structures, the developed analytical model provides sufficient accuracy in the calculations of spatial distributions in the main portion of the i-region
 and of the p-i-n structure overall  characteristics. The effects associated with the edge recombination and smearing of
 the p-i and i-n junctions are evaluated.

 The obtained results appear to be useful for the realization and optimization of GL- and MGL -based injection terahertz and infrared lasers.

\section*{Acknowledgments}
The authors are grateful to M. S. Shur and V. Mitin for numerous fruitful discussions.
The work was financially supported in part by the Japan Science and Technology Agency, CREST 
and by the Japan Society for Promotion of Science.


\begin{thebibliography}{99}

\bibitem{1}
A. H. Castro Neto, F. Guinea, N. M.R. Peres, K. S. Novoselov, and A. K. Geim,
REv. Mod Phys. {\bf 81}, 109 (2009).


\bibitem{2}
D. Farmer, Y.-M. Lin, A. Afzali-Ardakani, and P. Avouris,
Appl. Phys. Lett. {\bf 94}, 213106 (2009).

\bibitem{3}
E. C. Peters, E. J. H. Lee, M. Burghard, and Klaus Kern,
Appl. Phys. Lett. 97, 193102 (2010).

\bibitem{4} 
K. Yan,	Di Wu,	H. Peng,	Li Jin,	Q. Fu,	X. Bao, and 	
 Zh. Liu, Nat. Comm. {\bf 3}, 1280 (2012).

\bibitem{5} 
H. C. Cheng, R. J. Shiue, C. C. Tsai, W. H. Wang, Y. T. Chen,
ACS Nano  {\bf 5}, 2051 (2011).

\bibitem{6}
Yu Tianhua, C. Kim, C.-W. Liang, Yu Bin,
Electron Device Lett. {\bf 32}, 1050 (2011).

\bibitem{7}
V. V. Cheianov and V. I. Falko,
Phys. Rev. B {\bf 74}, 041403 (2006).

\bibitem{8}
B. Huard, J. A. Silpizio, N. Stander, K. Todd, B. Yang, and D. Goldhaber-Gordon,
Phys. Rev. Lett. {\bf 98}, 236803 (2007).

\bibitem{9}
J. R. Williams, L. DiCarlo, and C.M. Marcus,
Science {\bf 317}, 638 (2007).

\bibitem{10}
H-Y Chiu, V. Perebeinos, Y.-M. Lin, and P. Avouris,
Nano Lett. {\bf 10}, 4634 (2010). 

\bibitem{11}
M.~Ryzhii, V.~Ryzhii, T.~Otsuji, V.~Mitin, and M. S. Shur,
Phys. Rev. B {\bf 82}, 075419 (2010). 



\bibitem{12}
F. Xia, T. Mueller, Y.-M. Lin, A. Valdes-Garcia, and P. Avoris,
Nature Nanotech, {\bf 4}, 839 (2009).


\bibitem{13}V. Ryzhii, M. Ryzhii, V. Mitin, and T. Otsuji,
J. Appl. Phys. {\bf 107}, 054512 (2010). 

\bibitem{14}
T. Mueler, F. Xia, and P. Avouris,
Nature Photon. {\bf 4}, 297 (2010).

\bibitem{15}
M. Ryzhii, T. Otsuji, V. Mitin, and V.Ryzhii,
Jpn. J. Appl. Phys. {\bf 50}, 070117 (2011).

\bibitem{16}
N. M. Gabor, J. C. W. Song, Q. Ma, N. L. Nair, T. Taychatanapat, K. Watanabe, T. Taniguchi, L. S. Levitov, and P. Jarillo-Herrero
Science, {\bf 334}, 648 (2011).

\bibitem{17}
V. Ryzhii, N. Ryabova, M. Ryzhii, N. V. Baryshnikov, V. E. Karasik, V. Mitin.
and T. Otsuji,
Optoelectronics Rev. {\bf 20}, 15 (2012).


\bibitem{18}
D. Sun, G. Aivazian, A. M. Jones, J. Ross, W. Yao, D. Cobden, X. Xu, Nature Nanotechnology  
{\bf 7}, 114 (2012)

\bibitem{19}
M. Freitag, T. Low, and P. Avouris,
Nano Lett. {\bf 13 }, 1644 (2013).


\bibitem{20}
V. Ryzhii, M. Ryzhii, M. S. Shur, and V. Mitin,
Appl. Phys. Express {\bf 2}, 034503 (2009).

\bibitem{21}
V. L. Semenenko, V. G. Leiman, A. V. Arsenin, V. Mitin, M. Ryzhii, T. Otsuji,and V. Ryzhii,
J. Appl. Phys. {\bf 113},  024503  (2013).




\bibitem{22}
V. Ryzhii, M.~Ryzhii, and T. Otsuji,
 J. Appl. Phys. {\bf 101}, 083114 (2007).

\bibitem{23}
V.~Ryzhii, M.~Ryzhii, A.~Satou, 
T.~Otsuji, A.~A.~Dubinov, and 
V.~Ya. Aleshkin,
J. Appl. Phys. {\bf 106}, 084507 (2009).
%




\bibitem{24}
V. Ryzhii, A. A. Dubinov, T. Otsuji, 
V. Mitin, and M. S. Shur,
J. Appl. Phys. {\bf 107}, 054505  (2010). 

\bibitem{25}
A. A. Dubinov, V. Ya. Aleshkin, V. Mitin, T. Otsuji, and  V. Ryzhii,
J. Phys.: Condens. Matter {\bf 23}, 145302 (2011).



\bibitem{26}
S. Boubanga-Tombet, S. Chan, A. Satou, T. Otsuji, and V. Ryzhii Phys. Rev. B
{\bf 85}, 035443 (2012).

\bibitem{27}
 T.Otsuji, S. Boubanga-Tombet, A. Satou, M.Ryzhii, and V.Ryzhii,
 J. Phys. D {\bf 45}, 303001 (2012).

\bibitem{28}
T. Li, L. Luo, M. Hupalo, J. Zhang, M. C. Tringides, J. Schmalian, and J. Wang,
 Phys. Rev. Lett. {\bf 108}, 167401 (2012).



\bibitem{29}
M.~Ryzhii and V.~Ryzhii,
Jpn. J. Appl. Phys. {\bf 46}, L151 (2007).

\bibitem{30}
V. Ryzhii, M. Ryzhii, V. Mitin, and T. Otsuji,
J. Appl.Phys. {\bf 110}, 094503 (2011). 

\bibitem{31}
V.~Ryzhii, M.~Ryzhii, V.~Mitin, A.~Satou, and T.~Otsuji,
Jpn. J. Appl. Phys. {\bf 50}, 094001 (2011).



\bibitem{32}
D. Svintsov, V. Vyurkov, S. O. Yurchenko, T. Otsuji, and V. Ryzhii,
J. Appl. Phys. {\bf 111}, 083715 (2012).

\bibitem{33}
F. Rana, P. A. George, J. H. Strait, J. Dawlaty, S. Shivaraman, M. Chandrashekhar, and M. G. Spencer 
Phys. Rev. B 79, 115447 (2009).
\bibitem{34}
V. Ryzhii, M. Ryzhii, and T. Otsuji, Phys. Stat. Sol. (a) {\bf 205}, 1527 (2008).

\bibitem{35}
Ali H. Nayfeh {\it Perturbation Methods} (WILEY-VCH, Weinheim, 2004).

\bibitem{36}
R. S. Shishir and D. K. Ferry,
J. Phys. Condens. Matter {\bf 21}, 344201 (2009).




\bibitem{37}
M. S. Foster and I. Aleiner, Phys. Rev. B {\bf 79}, 085415  (2009).


\bibitem{38} L. E. Golub, S. A. Tarasenko, M. V. Entin, and L. I. Magarill, Phys. Rev. B 
{\bf 84}, 195408 (2011).

\bibitem{39}
A. Satou, V. Ryzhii, Y. Kurita, and T. Otsuji, J. Appl. Phys. {\bf  113}, 143108 (2013).


\bibitem{40} E. Kioupakis, P. Rinke, K. T. Delaney, and C. G. Van de Walle,
Appl. Phys. Lett. {\bf 98}, 161107 (2011).


\bibitem{41}
P. A. George, J. Strait, J. Davlaty, S. Shivaraman, M. Chandrashekhar, F. Rana, and M. G. Spencer,
Nano Lett. {vol. 8}, 4248 (2009). 



\bibitem{42}
T. Winzer and E. Malic, Phys. Rev B, {\bf 85}, 241404 (2012).

\bibitem{43}
F. Vasko and V. Mitin, Phys. Rev. B {\bf 84}, 155445 (2011). 

\bibitem{44} 
V. Vasko and V. Ryzhii, Phys. Rev. B {\bf 77}, 195433 (2008). 

\bibitem{45}
V. Ryzhii, M. Ryzhii, and T. Otsuji, Appl. Phys. Lett. {\bf 99}, 173504 (2001).

\bibitem{46}
O. G. Balev, F. T. Vasko, and V. Ryzhii,
Phys. Rev. B {\bf 79}, 165432 (2009).
\bibitem{47}
O. G. Balev and F. T. Vasko,
Phys. Rev. B {\bf 107}, 124312 (2010).
\end{thebibliography}
\end{document}